\documentstyle[12pt]{article}
\newcommand{\be}{\begin{equation}}
\newcommand{\ee}{\end{equation}}
\newcommand{\bea}{\begin{eqnarray}}
\newcommand{\eea}{\end{eqnarray}}
\newcommand{\ba}{\begin{array}}
\newcommand{\ea}{\end{array}}

\newcommand{\norsl}{\normalsize\sl}
\newcommand{\norsc}{\normalsize\sc}
\begin{document}

%-------------------- Title page ----------------------------------

\begin{titlepage}

\title{ 
\vskip -3cm
{\normalsize
\begin{flushright}
KUCP-101   \\
September, 1996
\end{flushright}
}
\vskip 3cm
Feigin-Fuchs Representations for Nonequivalent Algebras 
                of $N=4$ Superconformal Symmetery}

\author{
\vspace*{1cm}\\
\norsc  Satoshi MATSUDA
         \thanks{e-mail address:
           matsuda@phys.h.kyoto-u.ac.jp}
%\\
%\norsc
         \thanks{Work supported in part by
          the {\it Monbusho} Grant-in-Aid for Scientific Research
          on Priority Areas 231 ``Infinite Analysis", 
          No.08211227.}\\ 
\vspace{-5mm}\\
\norsl  Department of Fundamental Sciences, FIHS\\
\norsl  Kyoto University,  Kyoto 606-01, JAPAN\\
\vspace{2mm}
\\
\norsc  Yukitaka ISHIMOTO
%\thanks{
%                                           }
%\\
         \thanks{e-mail address:
             ishimoto@phys.h.kyoto-u.ac.jp}\\
\vspace{-5mm}\\
\norsl  Graduate School of Human and Environmental Studies\\
\norsl  Kyoto University, Kyoto 606-01, JAPAN}

\date{}
\maketitle

\vspace{1cm}

\begin{abstract}
{\normalsize
%\noindent
The $N=4$ SU(2)$_k$ superconformal algebra has 
the global automorphism of SO(4) $\approx$ SU(2)$\times$SU(2) 
with the {\it left} factor as the Kac-Moody gauge symmetry.
As a consequence, an infinite set of independent algebras labeled by 
$\rho$ corresponding to the conjugate classes of 
the {\it outer} automorphism group SO(4)/SU(2)=SU(2) 
are obtained \`a la Schwimmer and Seiberg. 
We construct Feigin-Fuchs representations 
with the $\rho$ parameter embedded
for the infinite set of the $N=4$ nonequivalent algebras. 
In our construction 
the extended global SU(2) algebras labeled by $\rho$  
are  self-consistently represented by fermion fields 
with appropriate boundary conditions.

                        }
\end{abstract}

\begin{picture}(5,2)(-300,-615)
\put(2.3,-110){}
\put(2.3,-125){}
\put(2.3,-140){}
\put(2.3,-155){}
\end{picture}  

\thispagestyle{empty}
\end{titlepage}

\setcounter{page}{1}
\baselineskip 24pt
%----------------------- Text -----------------------------------
Two dimensional conformal and superconformal field theories 
have become the fundamental subject for study 
in pursuing superstring theories in particle physics 
or two dimensional critical phenomena in statistical physics.
The underlying superconformal algebras with $N$=0,1,2,3 and 4 
have been much studied, and their representation theories 
have been investigated to a great extent.

In particular it is well recognized nowadays that 
the so-called Feigin-Fuchs (FF)
representations (or the Coulomb-gas representations)~\cite{FF} 
are very important and almost inevitably required tools for 
investigating the representation theories of the conformal 
and superconformal algebras.
By now we have established the FF representations of the 
superconformal algebras with higher number of supercharges
~\cite{DotseF,KatoM1/2,KatoM3,KatoMR,Miki}, 
up to $N$=4~\cite{Mat1,Mat2,ItoMP}. 

On the other hand, the spectral flows resulting from 
the {\it inner} automorphisms of 
the conformal and superconformal algebras 
with $N$=2,3 and 4 were first recognized 
by Schwimmer and Seiberg~\cite{SchwS}, 
and their remarkable implications on the representation theories of 
the algebras have been discussed by many people
~\cite{EguchiT3,DefeverST,PetersenT}.

In their same paper~\cite{SchwS} was 
presented the mechanism through which the truly different types of 
algebras arise for each $N$. 
In general, a given algebra has a global automorphism group $G$. 
Then, the different types of algebras are obtained by imposing 
boundary conditions on the generators of the given algebra
and are labeled by the conjugate classes of $G$. 
However, some of the {\it twists} introduced in this manner 
can be removed by operating the local gauge transformations 
on the algebras which are reflected 
by the presence of the Kac-Moody subalgebras. 
The truly independent algebras are then labeled by those 
conjugate classes of the global automorphisms 
not contained in the local automorphisms.
In other words, the independent algebras correspond to the 
conjugate classes of the {\it outer} automorphism group of the algebra, 
while the {\it inner} automorphisms give equivalent algebras.

In the present paper we shall study 
the $N=4$ SU(2)$_k$ superconformal algebra 
from this algebraic point of view. 
The $N=4$ SU(2)$_k$ superconformal algebra has the global SU(2) 
as well as the local SU(2) Kac-Moody symmetries embedded. 
Consequently the algebra exhibits the global automorphism 
group structure of SO(4)$\approx$SU(2)$\times$SU(2), 
resulting in the infinite set of algebras through periodicity 
conditions imposed on the generators. 
Removing some of the {\it twists} by use of 
the local SU(2) gauge transformations, 
the truly independent algebras are obtained which are 
labeled by $\rho$ corresponding to the conjugate classes of 
the {\it outer} automorphism group SO(4)/SU(2)=SU(2).

In order to make the paper self-contained and 
also to fix our notations, 
we shall first 
summarise the known results which are relevant to our study. 
Then, we shall construct 
the FF representations of the truly 
independent algebras being labeled by the $\rho$ parameter. 
The representation theories along this line has not been fully 
investigated so far.
The attempt to construct unitary representations of the 
$\rho$-extended  
algebras has been challenged~\cite{Yu}, 
but has remained to be unsuccessful. 
Our construction of the FF representations allow one to study 
not only  the unitary, but also the nonunitary representations of 
the infinite set of the 
$\rho$-extended algebras.

The $N$=4 SU(2)$_k$ superconformal algebra is defined by the 
form of the operator product expansions (OPE) among operators 
given by the energy-momentum tensor $L(z)$, 
the SU(2)$_k$ local nonabelian generators $T^i(z)$, and 
the iso-doublet and iso-antidoublet supercharges $G^a(z)$ 
and $\bar{G}_a(z)$~\cite{Mat1,Mat2,EguchiT12}:
\begin{eqnarray}
  L(z)L(w) &\sim& {3k\over (z-w)^4}+{2L(w)\over(z-w)^2}+
  {\partial_wL(w)\over z-w}\ ,                                \nonumber \\
  T^i(z)T^j(w) &\sim& {{1\over2}k\eta^{ij}\over(z-w)^2}+
      {{\rm i}\epsilon^{ijk}\eta_{kl}T^l(w)\over z-w}\ ,\ \ 
   \ L(z)T^i(w)\sim {T^i(w)\over (z-w)^2}+
        {\partial_wT^i(w)\over z-w}\ ,                         \nonumber \\
  T^i(z)G^a(w) &\sim& -{{1\over 2}{(\sigma^i)^a}_bG^b(w)\over z-w}\ ,\quad 
   \ \quad\qquad  T^i(z)\bar G_a(w)\sim 
      {{1\over 2}\bar G_b(w){(\sigma^i)^b}_a\over z-w}\ ,     \nonumber \\
  L(z)G^a(w) &\sim& {{3\over2}G^a(w)\over(z-w)^2}+
     {\partial_w G^a(w)\over z-w}\ ,\ \ \quad 
    \ L(z)\bar G_a(w)\sim{{3\over2}\bar G_a(w)\over(z-w)^2}+
         {\partial_w \bar G_a(w)\over z-w}\ ,                 \nonumber \\
  G^a(z)G^b(w) &\sim& 0\ ,\quad \quad 
  \hskip 3cm\quad \bar G_a(z)\bar G_b(w)\sim 0\ , \nonumber \\
  G^a(z)\bar G_b(w) &\sim& {4k{\delta^a}_b\over(z-w)^3}-
     {4{(\sigma^i)^a}_b{\eta_{ij}}T^j(w)\over(z-w)^2}-
     {2{(\sigma^i)^a}_b{\eta_{ij}}\partial_w T^j(w)\over z-w}+
     {2{\delta^a}_bL(w)\over z-w}\ , 
  \label{OPE}
\end{eqnarray}
where $i=0,\pm$ denote SU(2) triplets in the diagonal basis, while
the superscripts (subscripts) $a=1,2$ label SU(2) doublet (antidoublet) 
representations.
The symmetric tensors $\eta^{ij}=\eta^{ji}=\eta_{ij}$ 
are defined in the diagonal basis as 
$\eta^{+-}=\eta^{00}=1$, while the antisymmetric tensors 
$\epsilon^{ijk}=-\epsilon^{jik}=-\epsilon^{ikj}$ are 
similarly defined as $\epsilon^{+-0}=-{\rm i}$, etc., 
and otherwise zero.
The Pauli matrices are given by 
$\sigma^{\pm}=(\sigma^1\pm{\rm i}\sigma^2)/\sqrt 2, \sigma^0=\sigma^3$. 
The corresponding notations 
in terms of the isospin raising and lowering by a half unit
for the iso-doublet and iso-antidoublet fermionic operators 
are given by 
\bea
\Bigl(G^a(z)\Bigr) &=&\pmatrix{
                G^1(z) \cr
                G^2(z) \cr
                }
       =\pmatrix{
                G^-(z) \cr
                G^+(z) \cr
                }\ ,     \nonumber\\
\Bigl(\bar G_a(z)\Bigr) &=& \Bigl(\bar G_1(z), \bar G_2(z) \Bigr)
                         =\Bigl(\bar G_+(z), \bar G_-(z)   \Bigr)\ .
\label{DoubG}
\eea
The symmetric delta function ${\delta^a}_b={\delta^b}_a$ 
has the standard meaning with 
${\delta^-}_+={\delta^1}_1=1,\, {\delta^+}_+={\delta^2}_1=0, etc.$. 

The $N$=4 algebra of Eq.(\ref{OPE}) has the global SU(2) symmetry, 
whose zero-mode generators we hereby denote as 
${\bf S}_0=(S^+_0, S^-_0, S^0_0)$. 
The generators ${\bf S}_0$ are not included in the $N$=4 algebra, but 
are introduced here to classify the $N$=4 states:
\bea 
  [S^i_0, S^j_0] &=& {\rm i}\epsilon^{ijk}\eta_{kl}S^l_0\ , 
\nonumber\\~ 
  [S^i_0, \hat{G}^a(z)] &=& -{1\over 2}{(\sigma^i)^a}_b\hat{G}^b(z)\ , 
\quad 
  [S^i_0, \bar{\hat{G}}_a(z)] = {1\over 2}\bar{\hat{G}}_b(z)
  {(\sigma^i)^b}_a\ ,\nonumber\\~ 
  [S^i_0, T^j(z)] &=& [S^i_0, L(z)]=0\ ,
\label{defglob}
\eea
where the doublet and antidoublet combinations of 
supercharges, $\hat{G}^a$ and $\bar{\hat{G}_a}\  (a=1,2)$, 
under the global SU(2) symmetry are given by~\cite{MU12}
\bea
\Bigl(\hat{G}^a(z)\Bigr) &=& \pmatrix{ 
                       {\bar G}_1(z) \cr
                       -G^2(z)      \cr
             }
            =\pmatrix{
                       {\bar G}_+(z) \cr
                       -G^+(z)      \cr
             }\ , \nonumber\\
\Bigl(\bar{\hat{G}}_a(z)\Bigr) &=& \Bigl(G^1(z), -{\bar G}_2(z) \Bigr)
                   =\Bigl( G^-(z), -{\bar G}_-(z) \Bigr)\ .
\label{globDoubG}
\eea

The automorphism group for the above $N$=4 algebra 
is SO(4)$\approx$SU(2)$\times$SU(2), therefore the conjugate classes 
are characterized by two rotation angles, $2\pi\eta$ and $2\pi\rho$.
The corresponding boundary conditions are given by 
\begin{eqnarray} 
T^i(z) &=& e^{-2{\rm i}\pi(i\eta)}T^i(e^{2{\rm i}\pi} z)\ ,\nonumber\\ 
G^a(z) &=& -e^{{\rm i}\pi(\rho-a\eta)}
                                  G^a(e^{2{\rm i}\pi}z)\ ,\nonumber\\ 
\bar G_a(z) &=& -e^{-{\rm i}\pi(\rho+a\eta)}
                             \bar G_a(e^{2{\rm i}\pi}z)\ . 
\label{BoundC1}
\end{eqnarray}
where we have used the following notation: 
$i\eta=(\pm\eta,0)$ for $i=(\pm,0)$ 
and $a\eta=\pm\eta$ for $a=\pm$.

The local symmetry is just the SU(2) gauge symmetry which 
is the {\it left} factor of SO(4)$\approx$SU(2)$\times$SU(2).
We have therefore the {\it inner} automorphism 
specified by the local parameter $\alpha(z)$:
\bea
L(z) &\rightarrow& L(z)+{\rm i}{d\alpha(z)\over dz}T^0(z)
-{k\over 4}\left({d\alpha(z)\over dz}\right)^2\ , \nonumber \\
T^0(z) &\rightarrow& T^0(z)+{\rm i}{k\over 2}{d\alpha(z)\over dz}\ ,\qquad
T^{\pm}(z) \rightarrow e^{\pm{\rm i}\alpha(z)}T^{\pm}(z)\ , \nonumber \\
G^{\mp}(z) &\rightarrow& e^{\mp{\rm i}{\alpha(z)\over 2}}G^{\mp}(z)\ ,\quad
\qquad\bar G_{\pm}(z) \rightarrow 
e^{\pm{\rm i}{\alpha(z)\over 2}}\bar G_{\pm}(z)\ . 
\label{Auto}
\eea
One can therefore gauge away the $\eta$ phase 
by choosing $\alpha(z)={\rm i}\eta\log z$~\cite{SchwS}.

Since the algebra for each choice of $\eta$ is 
equivalent to each other 
through the inner automorphism mentioned above, 
we may only consider a typical value of $\eta$ like 
$\eta=0$ or $\eta=1$.
The value $\eta=0$ in Eq.(\ref{BoundC1}) corresponds 
to the Ramond (R) sector, 
while that of $\eta=1$ to the Neveu-Schwarz (NS) one.
In the following we shall consider the NS case for simplicity,
unless stated otherwise. 

As the result of taking $\eta=1$ 
we end up with the simpler periodicity conditions 
specified by $\rho$ only:
\bea 
T^{\pm}(z) &=& T^{\pm}(e^{2{\rm i}\pi} z)\ ,\nonumber\\  
G^{\mp}(z) &=& e^{{\rm i}\pi\rho}
G^{\mp}(e^{2{\rm i}\pi}z)\ ,\nonumber\\ 
\bar G_{\pm}(z) &=& e^{-{\rm i}\pi\rho}
\bar G_{\pm}(e^{2{\rm i}\pi}z)\ . 
\label{BoundC2}
\eea
Corresponding to the boundary conditions Eq.(\ref{BoundC2}), 
we have the following Fourier mode expansions for the generators 
of the NS sector:
\bea
L(z) &=& \sum_{n\in Z} L_n z^{-n-2}\ ,  \nonumber\\
T^i(z)&=&\sum_{n\in Z} T^i_n z^{-n-1}\ ,\nonumber\\
G^a(z)&=&\sum_{n\in Z} G^a_{n+{1\over 2}(1+\rho)} 
z^{-n-{1\over 2}(1+\rho)-{3\over 2}}\ , 
\nonumber\\
\bar G_a(z)&=& \sum_{n\in Z} \bar G_{a,n+{1\over 2}(1-\rho)}
z^{-n-{1\over 2}(1-\rho)-{3\over 2}}=
\sum_{n\in Z} \bar G_{a,n-{1\over 2}(1+\rho)}
z^{-n+{1\over 2}(1+\rho) -{3\over 2}}\ .
\label{modeExp2}
\eea

Thus we conclude that one can gauge away the $\eta$ phase
but {\it not} the $\rho$ phase, 
which implies that all algebras which differ by the value of 
the parameter $\eta$ are equivalent, whereas 
the remaining infinite set of the algebras labeled by the $\rho$ parameter 
are all nonequivalent to each other.

For the convenience of our later use, we shall give here 
in terms of the Fourier components 
the infinite set of nonequivalent 
$N$=4 SU(2)$_k$ superconformal algebras 
which are labeled by 
the continuous parameter $0\le\rho<2$: 
\begin{eqnarray}
  [L_m,L_n] &=& (m-n)L_{m+n}+{k\over 2}m(m^2-1)\delta_{m+n,0}\ , \nonumber\\~
  [T^i_m,T^j_n] &=& {\rm i}\epsilon^{ijk}\eta_{kl} T^l_{m+n}
           +\eta^{ij}{k\over 2}m\delta_{m+n,0}\ , \qquad        
  [L_m,T^i_n]=-nT^i_{m+n}\ ,                          \nonumber\\~
  [T^i_m,G^a_{n+{1\over 2}(1+\rho)}] &=& 
  -{1\over 2}{(\sigma^i)^a}_b G^b_{m+n+{1\over 2}(1+\rho)}\ , \nonumber\\~
  [T^i_m,{\bar G}_{a,n-{1\over 2}(1+\rho)}] &=&
  {1\over 2}{\bar G}_{b,m+n-{1\over 2}(1+\rho)}
        {(\sigma ^i)^b}_a\ ,                          \nonumber \\~
  [L_m,G^a_{n+{1\over 2}(1+\rho)}] &=&
  \left({1\over 2}m-n-{1\over 2}(1+\rho)\right)G^a_{m+n+{1\over 2}(1+\rho)}\ ,
  \nonumber\\~ 
  [L_m,\bar G_{a,n-{1\over 2}(1+\rho)}]&=&
  \left({1\over 2}m-n+{1\over 2}(1+\rho)\right)
  \bar G_{a,m+n-{1\over 2}(1+\rho)}\ ,    \nonumber \\~
  \{G^a_{m+{1\over 2}(1+\rho)},G^b_{n+{1\over 2}(1+\rho)}\} &=&0\ ,
  \qquad\qquad\quad
  \{\bar G_{a,m-{1\over 2}(1+\rho)}, \bar G_{b,n-{1\over 2}(1+\rho)}\}=0\ , 
  \nonumber\\~
  \{G^a_{m+{1\over 2}(1+\rho)},\bar G_{b,n-{1\over 2}(1+\rho)}\} 
  &=& 2{\delta^a}_b L_{m+n}
  -2(m-n+\rho+1){(\sigma^i)^a}_b \eta_{ij}T^j_{m+n} \nonumber\\~
    &&\qquad
  +{\delta^a}_b{k\over 2}\left(4\left(m+{1\over 2}(1+\rho)\right)^2-1\right)
     \delta_{m+n,0}\ , 
       \label{Fourier}
\end{eqnarray}
where $m$ and $n$ run over integers.
The modings for the extended SU(2) global generators 
${\bf S}_\rho=(S^+_\rho,S^-_{-\rho},S^0_0)$ 
are now given so that we have 
\bea
 [S^+_\rho,S^-_{-\rho}]&=&S^0_0+{\rho\over 2}\ ,
 \qquad\qquad\qquad\quad
 [S^0_0,S^{\pm}_{\pm\rho}]=\pm S^{\pm}_{\pm \rho}\ , \nonumber\\~
 [S^+_\rho,G^a_{n+{1\over 2}(1+\rho)}]&=&0\ ,\qquad\qquad\qquad
 [S^-_{-\rho},\bar G_{a,n+{1\over 2}(1-\rho)}]=0\ , \nonumber\\~
 [S^+_\rho,\bar G_{a,n+{1\over 2}(1-\rho)}] &=& {1\over \sqrt 2}
 \epsilon_{ab}G^b_{n+{1\over 2}(1+\rho)}\ ,\qquad
 [S^-_{-\rho}, G^a_{n+{1\over 2}(1+\rho)}] ={1\over \sqrt 2}
 \epsilon^{ab}\bar G_{b,n+{1\over 2}(1-\rho)}\ ,\nonumber\\~
 [S^0_0,G^a_{n+{1\over 2}(1+\rho)}]&=&
 {1\over 2}G^a_{n+{1\over 2}(1+\rho)}\ ,\qquad
 [S^0_0,\bar G_{a,n+{1\over 2}(1-\rho)}]
 =-{1\over 2}\bar G_{a,n+{1\over 2}(1-\rho)}\ ,\nonumber\\~
 [L_n,S^{\pm}_{\pm\rho}]&=&\mp\rho S^{\pm}_{n\pm\rho}\ ,\nonumber\\~
 [L_n,S^0_0]&=&[T^i_n,S^{\pm}_{\pm\rho}]=[T^i_n,S^0_0]=0\ ,
\label{Mode}
\eea
where the antisymmetric tensors are defined as 
$\epsilon_{12}=-\epsilon_{21}
=-\epsilon^{12}=\epsilon^{21}=1$, otherwise zero.
The c-number term $\rho/2$ 
in the right hand side of the first equality signals 
the presence of an anomaly when $\rho\not=0$.
Let us notice that, as this anomaly and 
the equality before the last 
in Eq.(\ref{Mode}) show, the global SU(2)$\times$SU(2) symmetry 
is broken down to SU(2)$\times$U(1) when $\rho\not=0$.
By the way we point out 
that the presence of the anomaly term just mentioned 
was overlooked in the paper by Yu~\cite{Yu}.

Here we take the raising operators to be 
\bea
L_n\quad &&(n>0)\ ,\nonumber\\
T^i_n\quad &&(n>0 \quad {\rm or}\quad  i=+\  {\rm and}\  n=0)\ ,\nonumber\\
G^a_{n+{1\over 2}(1+\rho)}\quad &&( n\ge 0 )\ ,\nonumber\\
\bar G_{a,n+{1\over 2}(1-\rho)}\quad &&( n\ge 0 )\ ,\nonumber\\
\delta_{\rho,0}S^+_\rho\quad &&\ ,
\label{RaisingOp2}
\eea
the Cartan subalgebra to be $\{L_0,T^0_0,k;S^0_0\}$, and 
the lowering operators to be the remaining generators. 
Then we define a highest weight representation (hwrep) of the algebras 
Eqs.(\ref{Fourier}) and (\ref{Mode}) to be one containing 
a highest weight state (hws) vector 
$|h,j;s\big>$ such that 
\bea
L_0|h,j;s\big>&=&h|h,j;s\big>\ ,\nonumber\\
T^0_0|h,j;s\big>&=&j|h,j;s\big>\ ,\nonumber\\
S^0_0|h,j;s\big>&=&s|h,j,;s\big>\ ,
\label{hws}
\eea
and
\be
X_+|h,j;s\big>=0\ ,
\label{Rhws}
\ee
for all raising operators $X_+$.

Now we present our construction of the FF representations for 
the infinite set of 
the $N$=4 SU(2)$_k$ algebras with $\rho$.
We use four real bosons $\varphi_\alpha(z)\ \ (\alpha=1,2,3,4)$, and 
four real fermions forming a pair of 
complex fermion doublet $\gamma^a(z)$ and 
antidoublet $\bar\gamma_a(z)$\ \ $(a=1,2$ or $\pm)$ 
under SU(2)$_k$:
\bea
\Bigl(\gamma^a(z)\Bigr)&=&\pmatrix{
                     \gamma^1(z)   \cr
                     \gamma^2(z)   \cr
                     }
           =\pmatrix{
                     \gamma^-(z)   \cr
                     \gamma^+(z)   \cr
                     }\ ,                   \nonumber\\
\Bigl(\bar\gamma_a(z)\Bigr)&=&\Bigl(\bar\gamma_1(z),\bar\gamma_2(z) \Bigr)
           =\Bigl(\bar\gamma_+(z),\bar\gamma_-(z)  \Bigr)\ ,
\label{gammaD}
\eea
whose propagators are given by 
\bea
\big<\varphi_\alpha(z)\partial\varphi_\beta(w)\big>
=\big<\partial\varphi_\beta(w)\varphi_\alpha(z)\big>
&=& {\delta_{\alpha \beta}\over z-w}\ , \nonumber\\
\big<\bar\gamma_a(z)\gamma^b(w)\big>_\rho
=-\big<\gamma^b(w)\bar\gamma_a(z)\big>_\rho
&=&{{\delta_a}^b\over z-w}
\left(z\over w\right)^{\rho\over 2}\ .
\label{Propagator}
\eea
Here we note that in accordance with the periodicity conditions 
Eq.({\ref{BoundC2}) the above fermion pairs have been taken to satisfy 
the following boundary conditions 
\be
\gamma^a(z)=e^{{\rm i}\pi\rho}\gamma^a(e^{2{\rm i}\pi}z)\ ,
\qquad\quad
\bar\gamma_a(z)=e^{-{\rm i}\pi\rho}\bar\gamma_a(e^{2{\rm i}\pi}z)\ .
\label{BoundFermi}
\ee
As a result the gamma pairs have the Fourier mode expansions given by 
\bea
\gamma^a(z)&=&\sum_{n\in Z}\gamma^a_{n+{1\over 2}(1+\rho)}
z^{-n-{1\over 2}(1+\rho)-{1\over 2}}\ ,\nonumber\\
\bar\gamma_a(z)&=&\sum_{n\in Z}\bar\gamma_{a,n+{1\over 2}(1-\rho)}
z^{-n-{1\over 2}(1-\rho)-{1\over 2}}\ .
\label{ModeFermi}
\eea

Corresponding to the transformation properties 
Eqs.(\ref{DoubG}) and (\ref{globDoubG}) of the supercharges, 
the following rearranged pairs of the fermion fields~\cite{MU12} 
are considered to transform as 
a doublet $\hat\gamma^a(z)$ 
and an antidoublet $\bar{\hat\gamma}_a(z)\ \  (a=1,2)$ 
pairs under the global SU(2) symmetry:
\bea
\Bigl(\hat\gamma^a(z)\Bigr)&=&\pmatrix{
                  \bar\gamma_1(z)   \cr
                  -\gamma^2(z)            \cr
                  }
               =\pmatrix{
                  \bar\gamma_+(z)  \cr
                  -\gamma^+(z)           \cr
                  }\ ,             \nonumber\\
\Bigl(\bar{\hat\gamma}_a(z)\Bigr)&=&
           \Bigl(\gamma^1(z), -\bar\gamma_2(z) \Bigr)
           =\Bigl( \gamma^-(z), -\bar\gamma_-(z)   \Bigr)\ .
\label{globgammaD}
\eea

Here we shall define 
an extended normal-ordering operation 
for non-zero $\rho$ 
such that our formalism of the FF representations be given 
self-consistently 
for any value of  $\rho$.
The $\rho$ parameter is only relevant to fermion fields, 
so the following definition 
of normal-ordering is sufficient for our present purpose:
\bea
{}^\times_\times\bar\gamma_a(z)\gamma^b(w)^\times_\times
&\equiv& \bar\gamma_a(z)\gamma^b(w)
-\big<\bar\gamma_a(z)\gamma^b(w)\big>_{\rho,div}\ ,   \nonumber\\
{}^\times_\times\bar\gamma_a(z)
\partial\gamma^b(w)^\times_\times
&\equiv& \bar\gamma_a(z)\partial\gamma^b(w)
-\big<\bar\gamma_a(z)\partial\gamma^b(w)\big>_{\rho,div}\ ,   \nonumber\\
{}^\times_\times\partial\bar\gamma_a(z)
\gamma^b(w)^\times_\times
&\equiv& \partial\bar\gamma_a(z)\gamma^b(w)
-\big<\partial\bar\gamma_a(z)\gamma^b(w)\big>_{\rho,div}\ ,    \nonumber\\
{}^\times_\times\partial\bar\gamma_a(z)
\partial\gamma^b(w)^\times_\times
&\equiv& \partial\bar\gamma^a(z)\partial\gamma_b(w)
-\big<\partial\bar\gamma_a(z)\partial\gamma^b(w)\big>_{\rho,div}\ ,   
\label{OurNormalO}
\eea
where the notation $\big<\cdots\big>_{\rho,div}$ in a given expression 
$\big<\cdots\big>_{\rho}$
stands for the {\it divergent} contribution 
when the limit $w\to z$ is taken in that expression.
They are simply given by the following $\rho$-independent expressions:
\bea
\big<\bar\gamma_a(z)\gamma^b(w)\big>_{\rho,div}
&=&{{\delta_a}^b\over z-w}\ ,           \nonumber\\
\big<\bar\gamma_a(z)\partial\gamma^b(w)\big>_{\rho,div}
&=&{{\delta_a}^b\over (z-w)^2}\ ,        \nonumber\\
\big<\partial\bar\gamma_a(z)\gamma^b(w)\big>_{\rho,div}
&=&-{{\delta_a}^b\over (z-w)^2}\ ,       \nonumber\\
\big<\partial\bar\gamma_a(z)\partial\gamma^b(w)\big>_{\rho,div}
&=&-{2{\delta_a}^b\over (z-w)^3}\ .
\label{divergent}
\eea
The above definition of extended normal-ordering
satsifies required properties 
like the antisymmetry for fermion fields 
whose few examples are
\be
{}^\times_\times\bar\gamma_a(z)\gamma^b(w)^\times_\times
=-{}^\times_\times\gamma^b(w)\bar\gamma_a(z)^\times_\times\ ,\qquad
{}^\times_\times\bar\gamma_a(z)\partial\gamma^b(w)^\times_\times
=-{}^\times_\times\partial\gamma^b(w)\bar\gamma_a(z)^\times_\times\ ,
\label{antisymmetry}
\ee
such that 
a Wick contraction can be performed in a consistent manner.

In Eq.(\ref{OurNormalO}), note the difference 
from the usual definition of 
normal-ordering which is given 
for example by 
\bea
:\bar\gamma_a(z)\gamma^b(w):
\equiv \bar\gamma_a(z)\gamma^b(w)
-\big<\bar\gamma_a(z)\gamma^b(w)\big>_\rho \ ,
\label{NormalO}
\eea
where $:(\cdots):$ implies that 
in the product $(\cdots)$ of operators 
all creation operators 
stand to the left of all annihilation operators.
Note also that we naturally have 
\be
{}^\times_\times\gamma^a(z)\gamma^b(w)^\times_\times
=:\gamma^a(z)\gamma^b(w):\ ,\qquad
{}^\times_\times\bar\gamma_a(z)\bar\gamma_b(w)^\times_\times
=:\bar\gamma_a(z)\bar\gamma_b(w):\ ,
\ee
while 
the difference between the two definitions 
Eqs.(\ref{OurNormalO}) and (\ref{NormalO}) is just 
given by a finite $\rho$ dependent c-number term.
In particular we have 
\bea
{}^\times_\times\bar\gamma_a(z)\gamma^b(z)^\times_\times
&=&:\bar\gamma_a(z)\gamma^b(z):+{\rho\over 2z}{\delta_a}^b\ , \nonumber\\
{}^\times_\times\Bigl(\partial\bar\gamma_a(z)\gamma^b(z)
-\bar\gamma_a(z)\partial\gamma^b(z)\Bigr){}^\times_\times
&=&:\Bigl(\partial\bar\gamma_a(z)\gamma^b(z)
-\bar\gamma_a(z)\partial\gamma^b(z)\Bigr):
+{\rho^2\over 4z^2}{\delta_a}^b\ ,
\label{composite1}
\eea
when the limit $w\to z$ is taken. 
Multiplying a fermion field $\gamma^c(w)$ or $\bar\gamma_c(w)$ 
to the first equality in the above and 
performing the Wick contraction to the operator products 
on both sides, 
one can also prove that the equalities 
\bea
{}^\times_\times\bar\gamma_a(z)\gamma^b(z)\gamma^c(z)^\times_\times
&=&-{}^\times_\times\gamma^b(z)\bar\gamma_a(z)\gamma^c(z)^\times_\times
\nonumber\\
&=&:\bar\gamma_a(z)\gamma^b(z)\gamma^c(z):
+{\rho\over 2z}{\delta_a}^b\gamma^c(z)
-{\rho\over 2z}{\delta_a}^c\gamma^b(z)\ ,  \nonumber\\
{}^\times_\times\bar\gamma_a(z)\gamma^b(z)\bar\gamma_c(z)^\times_\times
&=&-{}^\times_\times\gamma^b(z)\bar\gamma_a(z)\bar\gamma_c(z)^\times_\times
\nonumber\\
&=&:\bar\gamma_a(z)\gamma^b(z)\bar\gamma_c(z):
+{\rho\over 2z}{\delta_a}^b\bar\gamma_c(z)
-{\rho\over 2z}{\delta^b}_c\bar\gamma_a(z)\ , 
\label{composite2}
\eea
hold in the limit $w\to z$. 

With our definition 
${}^\times_\times{(\cdots)}^\times_\times$ 
of normal-ordering 
one can perform the OPE calculations in a compact manner 
for any values of $0\le\rho<2$ 
as you would do in the usual Neveu-Schwarz case with $\rho=0$. 
One should only keep in mind that 
$:(\cdots):$ in the NS case with $\rho=0$ 
be replaced by 
${}^\times_\times(\cdots)^\times_\times$ everywhere 
in the course of the calculations,
so that one can easily keep track of the 
$\rho$ dependent extra terms which show up 
as the differences between the two definitions of normal-ordering.  
Only at the end of the calculations one may use 
the relations like 
Eqs.(\ref{composite1}) and (\ref{composite2}) given above  
to transform the obtained results with 
${}^\times_\times{(\cdots)}^\times_\times$ 
into more familiar expressions with 
$:(\cdots):$ 
having $\rho$-dependent extra terms added.           

Now we first consider the SU(2)$_{\hat k}$ Kac-Moody subalgebras 
with level $\hat k$. The generators are given in terms of 
the first three bosons by~\cite{Mat1,Nemes}
\bea
J^0(z)&=&{\rm i}\sqrt{\hat k\over 2}\partial\varphi_3(z)\ , \nonumber\\
J^{\pm}(z)&=&:{\rm i\over\sqrt 2}
\left(\sqrt{\hat k+2\over 2}\partial\varphi_1(z)\pm
{\rm i}\sqrt{\hat k\over 2}\partial\varphi_2(z)\right)
e^{{\pm\rm i}\sqrt{2\over \hat k}
\left(\varphi_3(z)-{\rm i}\varphi_2(z)\right)}:\ .
\label{Nemesh}
\eea
The corresponding contribution to the energy-momentum tensor is then 
given in the Sugawara form as~\cite{Mat1,Mat2} 
\be
{1\over\hat k+2}\sum_{i,j=\pm,0}:J^i(z)\eta_{ij}J^j(z):
=-{1\over 2}\sum_{\alpha=1}^3 :\Bigl(\partial\varphi_\alpha(z)\Bigr)^2:
+{\rm i}{\tau\over 2}\partial^2\varphi_1(z)\ ,
\label{Sugawara}
\ee
where 
\be
\tau\equiv\sqrt{2\over \hat k+2}\ .
\label{tau}
\ee

The total energy-momentum tensor $L(z)$ is obtained by adding 
the contribution from the fourth boson and from the fermion doublets 
to Eq.(\ref{Sugawara}):
\bea
L(z)&=&-{1\over 2}\sum_{\alpha=1}^4 :\Bigl(\partial\varphi_\alpha(z)\Bigr)^2:
+\left({\rm i}{\tau\over 2}\partial^2\varphi_1(z)
-{\rm i}\kappa\partial^2\varphi_4(z)\right)         \nonumber\\
&&\hskip3cm+{1\over 2}{}^\times_\times
\Bigl(\partial\bar\gamma(z)\cdot\gamma(z)
-\bar\gamma(z)\cdot\partial\gamma(z)\Bigr){}^\times_\times   \nonumber\\
&=&-{1\over 2}\sum_{\alpha=1}^4 :\Bigl(\partial\varphi_\alpha(z)\Bigr)^2:
+\left({\rm i}{\tau\over 2}\partial^2\varphi_1(z)
-{\rm i}\kappa\partial^2\varphi_4(z)\right)         \nonumber\\
&&\hskip3cm+{1\over 2}
\Biggl[:\Bigl(\partial\bar\gamma(z)\cdot\gamma(z)
-\bar\gamma(z)\cdot\partial\gamma(z)\Bigr):
+{\rho^2\over 2z^2}\Biggr]
\label{TotalEM}
\eea
with the parameter 
\be
\kappa\equiv{\rm i\over 2}k\tau\ ,\qquad\qquad k\equiv\hat k+1\ .
\ee
The last line of Eq.(\ref{TotalEM}) is obtained 
by use of the second identity 
in Eq.(\ref{composite1}).

We also define the total SU(2)$_k$ Kac-Moody currents 
$T^i(z)\ \ (i=\pm,0)\ $ with level $k=\hat k+1$ by adding the 
fermionic contribution to $J^i(z)$:
\bea
T^i(z)&=&J^i(z)+{1\over 2}
{}^\times_\times\bar\gamma(z)\sigma^i\gamma(z){}^\times_\times \nonumber\\
&=&J^i(z)+{1\over 2}
:\bar\gamma(z)\sigma^i\gamma(z):\ ,
\label{TotalKM}
\eea
where the first identiy in Eq.(\ref{composite1}) was used to get the 
last expression.

Finally the $N=4$ supercurrents $G^a(z)$ and 
$\bar G_a(z)\ \ (a=1,2\ \  {\rm or}\ \  \pm)$ 
in our Feigin-Fuchs representations are given by 
\bea
G^a(z)&=&\gamma^a(z){\rm i}\partial\varphi_4(z)-2\kappa\partial\gamma^a(z)
-{\rm i}\tau J^i(z)\eta_{ij}\Bigl(\sigma^j\gamma(z)\Bigr)^a \nonumber\\
&&\hskip4cm+\ {\rm i}\tau{}^\times_\times
\Bigl(\bar\gamma(z)\cdot\gamma(z)\Bigr)\gamma^a(z)^\times_\times
\nonumber\\
&=&\gamma^a(z){\rm i}\partial\varphi_4(z)-2\kappa\partial\gamma^a(z)
-{\rm i}\tau J^i(z)\eta_{ij}\Bigl(\sigma^j\gamma(z)\Bigr)^a
\nonumber\\
&&\hskip4cm +\ {\rm i}\tau\Biggl[
:\Bigl(\bar\gamma(z)\cdot\gamma(z)\Bigr)\gamma^a(z):
+{\rho\over 2z}\gamma^a(z)\Biggr]\ ,
\nonumber\\
\bar G_a(z)
&=&\bar\gamma_a(z){\rm i}\partial\varphi_4(z)-2\kappa\partial\bar\gamma_a(z)
+{\rm i}\tau J^i(z)\eta_{ij}\Bigl(\bar\gamma(z)\sigma^j\Bigr)_a \nonumber\\
&&\hskip4cm-\ {\rm i}\tau{}^\times_\times
\Bigl(\bar\gamma(z)\cdot\gamma(z)\Bigr)\bar\gamma_a(z)^\times_\times
\nonumber\\
&=&\bar\gamma_a(z){\rm i}\partial\varphi_4(z)-2\kappa\partial\bar\gamma_a(z)
+{\rm i}\tau J^i(z)\eta_{ij}\Bigl(\bar\gamma(z)\sigma^j\Bigr)_a
\nonumber\\
&&\hskip4cm -\ {\rm i}\tau\Biggl[
:\Bigl(\bar\gamma(z)\cdot\gamma(z)\Bigr)\bar\gamma_a(z):
+{\rho\over 2z}\bar\gamma_a(z)\Biggr]\ ,
\label{Supercurrents}
\eea
where Eq.(\ref{composite2}) was used to obtain the second expressions 
of $G^a(z)$ and $\bar G_a(z)$.

It is to be pointed out here that,
when $\rho=0$,  the Feigin-Fuchs 
representations of Eqs.(\ref{TotalEM}),(\ref{TotalKM}),(\ref{Supercurrents}) 
presented above are just reduced as they should be 
to the form of those 
first constructed~\cite{Mat1} by one (S.M.) of the present authors.

Now, the generators 
${\bf S}_\rho=(S^+_\rho,S^-_{-\rho},S^0_0)$ 
or $S^i_{i\rho}\ \ (i=\pm,0)$  
of the extended SU(2) global symmetry defined in Eq.(\ref{Mode}) 
can be constructed as 
\bea
s^i_{i\rho}(z)&\equiv&{1\over 2}{}^\times_\times
\bar{\hat\gamma}(z)\sigma^i\hat\gamma(z)^\times_\times\ ,
\nonumber\\
S^i_{i \rho}&\equiv&{1\over 2\pi i}\oint 
s^i_{i \rho}(z)z^{i\rho}\ dz\ ,   
\label{DefS}
\eea
where we note the following notation:\ $i\rho=(\rho,-\rho,0)$ 
for each $i=(+,-,0)$. 
To be more explicit, we have 
for ${\bf s}_\rho(z)=\left(s^+_\rho(z),s^-_{-\rho}(z),s^0_0(z)\right)$ 
or $s^i_{i\rho}(z)$ 
\bea
s^+_\rho(z)&=&-{1\over \sqrt 2}
{}^\times_\times\gamma^-(z)\gamma^+(z)^\times_\times
=-{1\over 2\sqrt 2}\ 
\epsilon_{ab}:\gamma^a(z)\gamma^b(z):\ ,    \nonumber\\
s^-_{-\rho}(z)&=&-{1\over\sqrt 2}
{}^\times_\times\bar\gamma_-(z)\bar\gamma_+(z)^\times_\times
=-{1\over 2\sqrt 2}\ 
\epsilon^{ab}:\bar\gamma_a(z)\bar\gamma_b(z):\ ,    \nonumber\\
s^0_0(z)&=&{1\over 2}{}^\times_\times\Bigl(
\gamma^-(z)\bar\gamma_+(z)-\bar\gamma_-(z)\gamma^+(z)
\Bigr){}^\times_\times
=-{1\over 2}:\bar\gamma(z)\cdot\gamma(z):-{\rho\over 2z}\ ,
\label{ExplicitS}
\eea
with the mode expansion of $\gamma^a(z)$ and 
$\bar\gamma_a(z)$ given by Eq.(\ref{ModeFermi}).

With this representation of Eq.(\ref{ExplicitS})
we actually generate the $\rho$-extended SU(2) local algebra 
\be
[S^i_{m+i\rho}, S^j_{n+j\rho}]
={\rm i}\epsilon^{ijk}\eta_{kl}S^k_{m+n+k\rho}
+\eta^{ij}{1\over2}(m+i\rho)\delta_{m+n,0}\ ,
\label{ModeS}
\ee
with the following mode expansion:
\be
s^i_{i\rho}(z)=\sum_{n\in Z}S^i_{n+i\rho}z^{-n-i\rho-1}\ .
\label{SmodeEx}
\ee
By putting $m=n=0$ in Eq.(\ref{ModeS}) 
we obtain the $\rho$-extended algebra of the SU(2) global symmetry 
given in Eq.(\ref{Mode}). 

With Eqs.(\ref{TotalEM}) and (\ref{ExplicitS}) 
we also have the OPE relation 
\be
L(z)s^i_{i\rho}(w)\sim\partial_w\left({1\over z-w}s^i_{i\rho}(w)\right)\ 
\qquad (i=+,-,0)\ ,
\label{OPELS}
\ee
which reproduces the last two commutation relations 
between $L_n$ and $S^i_{i\rho}$ in Eq.(\ref{Mode}) 
if it is transformed into Fourier modes.
Thus we have 
\be
[L_n, S^i_{i\rho}]=-i\rho S^i_{n+i\rho}\ 
\qquad (i=+,-,0)\ .
\label{ModeLS}
\ee

Now we can construct highest weight states (hws) explicitly 
in terms of the four bosons $ \varphi_\alpha(z)\ (\alpha=1,2,3,4)$ 
and four fermions $\gamma^a(z), \bar\gamma_a(z)\ (a=1,2)$.
The mode expansions for the latter are given by Eq.(\ref{ModeFermi}), 
whereas those for the former are given by 
\be
\varphi_\alpha(z)=q_\alpha-{\rm i}p_\alpha\log z
+{\rm i}\sum_{n\in Z,n\not=0}{\varphi_{\alpha,n}\over n}z^{-n}
\qquad(\alpha=1,2,3,4)\ .
\ee
The commutators for the Fourier modes are 
\bea
&&[\varphi_{\alpha,m},\varphi_{\beta,n}]=m\delta_{\alpha\beta}
\delta_{m+n,0}\ ,\qquad
[q_\alpha,p_\beta]={\rm i}\delta_{\alpha\beta}\ ,\nonumber\\
&&[\varphi_{\alpha,n},q_\beta]=[\varphi_{\alpha,n},p_\beta]=0\ ,
\nonumber\\
&&\{\gamma^a_{m+{1\over 2}(1+\rho)},\bar\gamma_{b,n-{1\over 2}(1+\rho)}\}
={\delta^a}_b \delta_{m+n,0}\ ,\nonumber\\
&&\{\gamma^a_{m+{1\over 2}(1+\rho)},\gamma^b_{n+{1\over 2}(1+\rho)}\}
=\{\bar\gamma_{a,m-{1\over 2}(1+\rho)},\bar\gamma_{b,n-{1\over 2}(1+\rho)}\}
=0\ .
\label{Commutator}
\eea

The ground state vacuum $|0\big>$ is now defined by 
\bea
\varphi_{\alpha,n}|0\big>=p_\alpha|0\big>=0
\quad\qquad\qquad&&(n>0)\ ,\nonumber\\
\gamma^a_{m+{1\over 2}(1+\rho)}|0\big>=
\bar\gamma_{a,m+{1\over 2}(1-\rho)}|0\big>=0\quad&&(m\ge0)\ .
\label{Vacuum}
\eea
With the vertex operators 
\bea
V(t,j,j_0;z)&=&:e^{{\rm i}t\varphi_4(z)}V_{j,j_0}(z):\ ,\nonumber\\
V_{j,j_0}(z)&=&:e^{-{\rm i}j\tau\varphi_1(z)}
e^{{\rm i}j_0\sqrt{2\over \hat k}
\bigl(\varphi_3(z)-{\rm i}\varphi_2(z)\bigr)}:\ ,
\label{Vertex}
\eea
which were first introduced 
by one (S.M.) of the present authors\cite{Mat1,Mat2}, 
a primary state with conformal dimension $h_\rho(t,j)$ and 
SU(2) spin $(j,j_0)$ is represented as 
\be
|h_\rho(t,j),j_0\big>\hspace{-1mm}\big>\sim V(t,j,j_0;z=0)|0\big>\ ,
\label{Primary}
\ee
where the conformal weight $h_\rho(t,j)$ is given by 
\bea
h_\rho(t,j)&\equiv& {t^2\over 2}+\kappa t+{\tau^2\over 2}j(j+1)
           +{\rho^2\over 4}  \nonumber\\
      &=&{1\over 2}(t+\kappa)^2+{\tau^2\over 2}(j+{1\over 2})^2
      +{\hat k\over 4}+{\rho^2\over 4}\ .
\label{Weight}
\eea
Note also that the following OPE relations hold\cite{Mat1,Mat2}:
\bea
J^0(z)V_{j,j_0}(w)&\sim&{j_0\over z-w}V_{j,j_0}(w)\ , \nonumber\\
{\sqrt 2}J^{\pm}(z)V_{j,j_0}(w)&\sim&
{-j\pm j_0\over z-w}V_{j,j_0\pm1}(w)\ .
\label{OPEJ}
\eea
Thus a hws vector $|h_\rho,j;s_\rho\big>$ is now given as 
\bea
\Big|h_\rho=h_\rho(t,j),j;s_\rho=-{\rho\over 2}\Big>
&\equiv&|h_\rho(t,j),j_0=j\big>\hspace{-1mm}\big>   \nonumber\\
&\sim&V(t,j,j_0=j;z=0)|0\big>\ .
\label{hwrep}
\eea

Here we have some remarks in order. 
First, as to the realization 
of the boundary conditions Eq.(\ref{BoundC2}) 
by our FF representations, we note that 
the bosonic part $J^i(z)$ of the SU(2)$_k$ Kac-Moody currents $T^i(z)$ 
satisfies the periodicity equations
\bea
J^0(z)&=&J^0(e^{2{\rm i}\pi}z)\ ,   \nonumber\\
J^{\pm}(z)&=&J^{\pm}(e^{2{\rm i}\pi}z)
e^{\mp2{\rm i}\pi\sqrt{2\over \hat k}(p_3-{\rm i}p_2)}\ ,
\label{BosonicBC}
\eea
as operator identities, while, 
as is clear from our vertex operator expressions Eq.(\ref{Vertex}), 
the momentum eigenvalues always satisfy 
\be
p_3-{\rm i}p_2=0
\ee
for any conformal states spanned on the primary state 
Eq.(\ref{Primary}) in our FF representation. 
Thus the boundary conditions Eq.(\ref{BoundC2}) 
is guaranteed to hold in a nontrivial manner 
for our FF representations. 

Secondly, 
the ``charge'' operator $\hat C$ considered 
by Kent and Riggs \cite{KentRiggs} 
is defined here as 
\be
\hat C\equiv 2S^0_0+\rho\ .
\ee
In particular we have for the hws 
\be 
\hat C|h_\rho,j;s_\rho\big>=0\ .
\ee

Lastly, 
the irreducible hwreps of the $\rho$-extended $N=4$ algebras Eq.(10) 
can be constructed as quotients of its Verma 
modules \cite{GoddardOlive}, 
which we denote by $V(h_\rho,j,k;\rho)$.
Here we define the ordering of numbers by writing 
$(p,r)<(q,s)$ if $p<q$ or both $p=q$ and $r<s$. 
Then, we have that if $x<y$, $n_x<n_y$ and 
$(p_x,r_x)<(q_y,s_y)$. 
With this notation of ordering, the Verma module is spanned by a 
basis of the states of the form
\bea
&&\Biggl(\prod_{i=1}^{i_0}\left(L_{n_i}\right)^{\ell_{n_i}}\Biggr)
\Biggl(\prod_{i=(\pm,0)}
\prod_{j_i=1}^{j_{i,0}}\left(T^i_{m_{j_i}}\right)^{t_{m_{j_i}}}\Biggr)
\Biggl(\prod_{a=\pm}
\prod_{k_a=1}^{k_{a,0}}\left(G^a_{p_{k_a}+{1\over 2}(1+\rho)}
\right)\Biggr)      \nonumber\\ 
&&\hskip5cm 
\times\Biggl(\prod_{b=\pm}\prod_{l_b=1}^{l_{b,0}}
\left(
\bar G_{b,q_{l_b}+{1\over 2}(1-\rho)}
\right)\Biggr)
|h_\rho,j;s_\rho\big>\ ,
\label{Basis}
\eea
where only lowering operators appear in the products, 
and 
the modes $n_i,m_{j_i},p_{k_a},q_{l_b}$ are negative integers while
the powers $\ell_{n_i}, t_{m_{j_i}}$ 
are positive integers.
Defining

\bea
N&\equiv&\sum_{i=1}^{i_0}(-n_i) \ell_{n_i}
+\sum_{i=(\pm,0)}\sum_{j_i=1}^{j_{i,0}}(-m_{j_i})t_{m_{j_i}}
+\sum_{a=\pm}\sum_{k_a=1}^{k_{a,0}}(-p_{k_a}-{1\over 2})
\nonumber\\
&&\hskip8cm
+\sum_{b=\pm}\sum_{l_b=1}^{l_{b,0}}(-q_{l_b}-{1\over 2})\ ,
\nonumber\\
K&\equiv&\sum_{i=(\pm,0)}\sum_{j_i=1}^{j_{i,0}}it_{m_{j_i}}
+\sum_{a=\pm}\sum_{k_a=1}^{k_{a,0}}a{1\over 2}
+\sum_{b=\pm}\sum_{l_b=1}^{l_{b,0}}b{1\over 2}   \nonumber\\
&=&\sum_{i=(\pm,0)}\sum_{j_i=1}^{j_{i,0}}it_{m_{j_i}}
+\sum_{a=\pm}a{1\over 2}{k_{a,0}}
+\sum_{b=\pm}b{1\over 2}{l_{b,0}}\ ,
\nonumber\\
C&\equiv&\sum_{a=\pm}\sum_{k_a=1}^{k_{a,0}}1
  +\sum_{b=\pm}\sum_{l_b=1}^{l_{b,0}}(-1)  \nonumber\\
 &=&\sum_{a=\pm}{k_{a,0}}
  -\sum_{b=\pm}{l_{b,0}}\ ,
\label{Eigenvalue}
\eea
we find that 
the state Eq.(\ref{Basis}) has $L_0$-eigenvalue $h_\rho+N-{1\over 2}C\rho$, 
$T^0_0$-eigenvalue $j+K$, and $\hat C$-eigenvalue $C$. 
The Verma module splits into simultaneous eigenspaces of 
$L_0, T^0_0$ and $\hat C$, whose splitting can be written as 
\be
V(h_\rho,j,k;\rho)={\bigoplus_{\tiny\begin{array}{c}
                            N\ge 0 \\
                            K\in {1\over 2}Z \\
                            C\in Z 
                            \end{array}}}
   V_{N,K,C}(h_\rho,j,k;\rho)
\label{Splitting}
\ee

In conclusion, we have considered  
the infinite number of nonequivalent algebras 
of $N=4$ SU(2)$_k$ superconformal symmetry labeled by 
the $\rho$ parameter which specifies 
the global boundary conditions for the generators of 
the $N=4$ algebra. 
We have presented the FF representations of the 
$\rho$-extended $N=4$ algebras in terms of four bosons and 
a pair of complex fermion doublets. 
The generators of the $\rho$-extended 
SU(2) global symmetry are explicitly constructed 
in terms of the fermion pairs. 
The formalism of the hwrep's of the $\rho$-extended 
$N=4$ SU(2)$_k$ superconformal algebras has been 
given in our FF representations with $\rho$.

\vspace{0.5cm}
One of the authors (S.M) would like to thank 
Professor Tsuneo Uematsu for discussions.
%%%%%%%%

%%%%%%%%%%%%%%%%%%%%%%%%%%%%%%%%%%%%%%%%%%%%%%%%%%%%%%%%%%%%%%%%%%
%%%%%%%%%%%%%%%%%%%%%%%%%%%%%%%%%%%%%%%%%%%%%%

%%%%%%%%%%%%%%%%%%

\end{document}